# Algorithmic decomposition for efficient multiple nuclear spin detection in diamond

*Hyunseok Oh[1]\*, Jiwon Yun[1]\*, M.H. Abobeih[2,3], Kyung-Hoon Jung[1], Kiho Kim[1], T.H. Taminiau[2,3], and Dohun Kim[1]\*\**

[1]*Department of Physics and Astronomy, and Institute of Applied Physics, Seoul National University, Seoul 08826, Korea*

[2]*QuTech, Delft University of Technology, PO Box 5046, 2600 GA Delft, The Netherlands*

[3]*Kavli Institute of Nanoscience Delft, Delft University of Technology, PO Box 5046, 2600 GA Delft, The Netherlands*

*\* These authors contributed equally to this work.*

*\*\*Corresponding author:* [dohunkim@snu.ac.kr](dohunkim@snu.ac.kr)

**Abstract**

Efficiently detecting and characterizing individual spins in solid-state hosts is an essential step to expand the fields of quantum sensing and quantum information processing. While selective detection and control of a few $^{13}$C nuclear spins in diamond have been demonstrated using the electron spin of nitrogen-vacancy (NV) centers, a reliable, efficient, and automatic characterization method is desired. Here, we develop an automated algorithmic method for decomposing spectral data to identify and characterize multiple nuclear spins in diamond. We demonstrate efficient nuclear spin identification and accurate reproduction of hyperfine interaction components for both virtual and experimental nuclear spectroscopy data. We conduct a systematic analysis of this methodology and discuss the range of hyperfine interaction components of each nuclear spin that the method can efficiently detect. The result demonstrates a systematic approach that automatically detects nuclear spins with the aid of computational methods, facilitating the future scalability of devices.

**Introduction**

Diamond nitrogen-vacancy (NV) centers have recently attracted much attention in magnetic resonance and related fields of research such as quantum metrology[1-5] and quantum information processing[6-9] because of their good quantum coherence. The spin states of NV centers show long coherence times[10,11] even at room temperature[12,13], can be manipulated by microwaves[14,15], can be connected optically over long distances[16], and can be operated in a wide range of temperatures through the measurement of optically detected magnetic resonance (ODMR) using spin-dependent fluorescence[17]. The NV center electron spin is also used as a sensitive probe of surrounding spins (most commonly used are $^{13}$C nuclear spins interacting with NV center electron spin by hyperfine coupling) as schematically shown in Fig. 1a. Owing to the long coherence times of these nuclear spins[18], detecting and manipulating naturally trapped nuclear spins in diamond through NV centers is emerging as a promising pathway to enable controlled quantum systems such as spin qubit registers[19-24], quantum networks[25,26], quantum simulation platforms[18,27], and quantum memories[28-30]. Because naturally occurring $^{13}$C nuclear spins are randomly located around an NV center electron spin[31,32], the efficient and automatic characterization of hyperfine interactions of individual nuclear spins is a prerequisite step towards building scalable quantum systems based on this platform.

Automatic methods that efficiently and objectively detect and characterize the underlying nuclear spins are required for further scalability of this platform. The detection of nuclear spins in diamond relies on the hyperfine coupling to NV center electron spin, and combined effects from each spin are reflected in the spectrum signal[31,32]. To identify each spin individually, the signal should be decomposed into multiple spectra, each resulting from a single nuclear spin. Therefore, as the spin system size increases, detecting individual nuclear spin signals from the observed magnetic resonance spectrum requires a more intricate analysis process. Applying for larger-scale spin systems, systematic and reliable approaches, currently lacking in the field of NV center-based nuclear spectrum analysis, are required.

In this study, we propose and demonstrate a computer-aided algorithm for analyzing nuclear spectra in diamond, resolving periodic dips automatically and determining hyperfine interaction tensor components of the nuclear spins weakly coupled to an NV center. Central to the algorithm, we adapted the Gaussian mixture model based on signal partitioning algorithms[33] to identify each dip from the signal, enabling analysis of the hyperfine interaction tensor components of each nuclear spin. We demonstrate systematic and automatic decomposition of both virtual and experimental dynamical decoupling-based nuclear spectroscopy data. We also show that the algorithm is capable of detecting the magnitude of hyperfine interaction tensor within the uncertainty of 20%, by comparing the results to those from a more complex and demanding multidimensional spectroscopy method applied to the same nuclear spins[23]. Our result provides a useful, fast and general tool for analyzing a wide variety of spectroscopy data; this is a key tool for expanding the number of available coherent resources in a system using hyperfine coupling to NV centers and nuclear spins.

**Materials and Methods**

*The characteristic signal*

In NV center-based nuclear magnetic resonance spectroscopy, dynamical decoupling pulse sequences are applied to detect a weak signature of individual nuclear spin signals. Periodic pulses are used to decouple the interaction of the NV center with a spin bath while amplifying specific nuclear Larmor precession resonant with the inter $\pi$-pulse period (Fig. 1b)[31]. In this study, we analyze Carr-Purcell-Meiboom-Gilbert (CPMG)[34,35] type dynamical decoupling sequences. We choose this sequence because it forms the basic work horse of spin detection[20,21,31,36], is of low experimental complexity and forms the starting point for more extensive characterization methods[23,37,38].

The interaction of the NV center with an individual nuclear spin is reflected in sudden dips in the measured electron coherence occurring periodically as a function of total phase evolution time. The analytic expression for the NV center coherence in the presence of nuclear spin is given by[31],

$$P_x = \frac{M+1}{2} \quad (1)$$

$$M = 1 - (1 - \hat{n}_0 \cdot \hat{n}_1)\sin^2 \frac{N\phi}{2} \quad (2)$$

where $P_x$ is the probability that the initial state is preserved, $N$ is the number of the repeated unit decoupling sequence, $\hat{n}_0$ and $\hat{n}_1$ are the rotation axes of the nuclear spin depending on the initial state of the electron spin with magnetic quantum number $m_s$=0, -1, respectively, $\phi$ is the net rotation angle around the axis $\hat{n}_i$ ($i$ = 0,1) following the relation $\cos\phi = \cos\alpha \cos\beta - m_z \sin\alpha \sin\beta$, and $1 - \hat{n}_0 \cdot \hat{n}_1 = m_x^2 \frac{(1-\cos\alpha)(1-\cos\beta)}{1+\cos\alpha\cos\beta - m_z\sin\alpha\sin\beta}$, where

$\alpha = \tilde{\omega}\tau$, $\beta = \omega_L\tau$, $m_z = (A+\omega_L)/\tilde{\omega}$, $m_x = B/\tilde{\omega}$, and $\tilde{\omega} = \sqrt{(A+\omega_L)^2 + B^2}$. For example, Fig. 1c shows numerically simulated spectroscopy data with a single nuclear spin near an NV center. The spectra show sharp dips in the coherence of the electron spin, with a position and amplitude determined by the parallel (perpendicular) hyperfine interaction tensor components $A$ ($B$).

In a typical dynamical decoupling experiment using a diamond chip with naturally abundant $^{13}$C nuclear spins, the spectroscopy signal shows a complicated spectrum stemming from each nuclear spin surrounding the electron spin superposed with a broad background signal. The random location of nuclear spins and the non-linearity of the signal makes stereotypical signal library construction impractical, and the peak detection and analysis should be performed for each given NV center and the uniquely associated nuclear spin bath[31,32]. These factors forbid direct application of conventional Nuclear Magnetic Resonance (NMR) spectrum analysis methods to NV center-based NMR signal processing.

The purpose of our computer-aided spectrum analysis is to accurately determine individual nuclear spins and obtain both longitudinal and transverse hyperfine parameters automatically. The signal analysis process can be divided into three steps: decomposition of a given signal into Gaussians using signal partitioning algorithm[33]; extraction of a single spin signal; and calculation of the hyperfine interaction of each spin. Each stage is fully automated and executed in sequential order, using user-defined parameters which are summarized in Table 1.

*Signal decomposition into Gaussians*

The first step of this method is to apply the signal partitioning algorithm to decompose the given CPMG-based dynamical decoupling signal into a combination of Gaussian functions. The algorithm first automatically divides the spectroscopy data into fragments, then each fragment is determined by a distinct splitting dip in the data. The data is divided so that nominally only one coherence dip exists in one fragment. All fragments are then decomposed into Gaussians based on Expectation-Maximization (EM) iterations. All Gaussian outputs with an amplitude smaller than the *threshold* are eliminated during the post-processing. MATLAB packages: *Signal Processing Toolbox*, *Bioinformatics Toolbox*, *Curve Fitting Toolbox*, and

*Parallel Computing Toolbox* are used for the implementation of the algorithm[33].

*Detection of single nuclear spins*

Resonance dips stemming from nuclear spins appear in the CPMG signal when the condition $\hat{n}_0 \cdot \hat{n}_1 = -1$ is met. In terms of phase evolution time in the $k^{th}$ fragment $\tau_k$, this condition can be approximated as

$$\tau_k = \frac{\pi}{\tilde{\omega} + \omega_L}\left[(2k-1) + (-1)^{k-1}\left(\frac{1}{2\sqrt{2}}\left(\frac{B}{\omega_L + A}\right)^2 - \frac{1}{4\sqrt{2}}\left(\frac{B}{\omega_L + A}\right)^4\right)\right] + O\left(\left(\frac{B}{\omega_L + A}\right)^6\right) \quad (3)$$

where $\omega_L$ is the Larmor frequency of the $^{13}$C nuclear spins, and $\tilde{\omega} = \sqrt{(\omega_L + A)^2 + B^2}$. This approximation can be derived from finding the intersection between the curves $\tan(\alpha/2)\tan(\beta/2) = 1/m_z$ and $\beta = \frac{\omega_L}{\tilde{\omega}}\alpha$ in the $\beta$ - $\alpha$ plot, using Taylor expansion with $\frac{B}{\omega_L + A}$, where $A(B)$ is the parallel(perpendicular) hyperfine interaction tensor component, $\alpha = \tilde{\omega}\tau$, $\beta = \omega_L \tau$, and $m_z = (A + \omega_L)/\tilde{\omega}$. When $B \ll \omega_L$, further approximation yields a linear $\tau_k$ vs. $k$ as follows:

$$\tau_k = \frac{(2k-1)\pi}{\tilde{\omega} + \omega_L}. \quad (4)$$

Therefore, the nuclear spin is identified by collecting the position of the $k^{th}$ coherence dip as a function of $k$, and fitting the result with a straight line. In this study, the dip positions from the signal decomposition process are partitioned into time windows, following the period $T = \frac{\pi}{\omega_L}$, and each dip position is marked with the relative time $\Delta\tau_k$ with respect to the center of the time window $\Delta\tau_k = \tau_k - \left(k - \frac{1}{2}\right)T$.

When collected, dip positions $\Delta\tau_k$ vs. $k$ form a fan diagram, and the CPMG line fit method composed of the dip grouping and a line fitting process is applied to the plot. With the constraint that all lines start from $(k, \Delta\tau_k) = (0.5, 0)$, satisfying Eq. (4), the objective of this process is to find a set of line configurations that explains the positions of all dips in the $\Delta\tau - k$ plot. We use a sequential process to reduce the computational load during the process. First, a set of line candidate lists is built based on the individual dip positions. For each dip located at $(k_0, \Delta\tau_0)$ in the $\Delta\tau - k$ plot, a list of line candidates starting from $(k, \Delta\tau) = (0.5, 0)$ and having slope $\dfrac{d\tau}{dk} = \dfrac{\Delta\tau_0}{(k_0 - 0.5)}$ is generated.

Second, for all line candidates in the list, we find the nearest dips to the line for each $k$, and calculate the squared distance of $\tau$ between the dips and the line. The mean squared distance of each line was calculated using $\overline{d^2} = \dfrac{\sum_{k=1}^{k_{max}} \min\left(\Delta\tau_{line}(k) - \Delta\tau_{dip}(k)_{min}, d_{max}\right)^2}{N}$, where

$\Delta\tau_{line}(k) = \left(\dfrac{d\tau}{dk}(k - 0.5) + 0.5\right)$ is the position of the line at period number k, and $\Delta\tau_{dip}(k)_{min}$ is the position of the dip with the closest distance to the line in consideration. To prevent picking dips too far from the line, a maximum value $d_{max}$ of the distance between the line and a dip is set. The line candidate having the least mean distance value among the $k$ values is considered as the best fit, and the dip positions used in the calculation are grouped as signals of the same nuclear spin. Third, the examined line candidate as well as a determined group of dip positions are removed from the $\Delta\tau - k$ plot. This process is repeated until all the points in the $\Delta\tau - k$ plot are exhausted.

During the line selection process where $k_{max}$ is the maximum time window index within the data aimed to be analyzed, lines with slopes larger than $T/2k_{max}$ (smaller than $-T/2k_{max}$) have $k$ values with 0(2) dips within them. Furthermore, all of the following dips within time windows with $k$ values larger than the window with 0(2) dips show the discrepancy between its time window index $k$ and dip numbering index. To solve this discrepancy, additional dips with coordinates $(k \pm m, \Delta\tau_k \mp mT)$, $m < M$, $m = $ integer, where $M$ is the number of additional layers of dips, are added for each dip in position $(k, \Delta\tau_k)$ in the $\Delta\tau - k$ plot. This procedure expands the period window to $(2M-1)T$, and the range of slope that can be fit into a straight line is expanded to $(-(2M-1)T/2k_{max}, (2M-1)T/2k_{max})$.

*Fitting and post selection*

The full-width-half-maximum (FWHM) of each dip grouped in the CPMG line fit and the slope obtained in the CPMG line fit are used as the two constraints determining ($A$, $B$) of each nuclear spin. For the first constraint, the envelope of the CPMG signal represented by a Lorentzian is used,

$$1 - \hat{n}_0 \cdot \hat{n}_1 = \frac{2}{1 + \left(\frac{(\tilde{\omega}+\omega_L)\tilde{\omega}(\tau - \tau_k)}{B}\right)^2} \tag{5}$$

which is valid for $\tau - \tau_k \ll \frac{B}{(\tilde{\omega}+\omega_L)\tilde{\omega}}$. From the Gaussian peak decomposition, we gain the standard deviation $\sigma$ of the best Gaussian fit to the dip, giving the FWHM of the fit as $2\sigma\sqrt{2\ln 2}$. Using this value as the FWHM of the Lorentzian envelope, the first relation between $A$, $B$, and $\sigma$ can be derived as follows:

$$\sigma = \frac{1}{\sqrt{2\ln 2}} \frac{B}{(\tilde{\omega}+\omega_L)\tilde{\omega}}. \tag{6}$$

The second constraint comes from the slope obtained during the CPMG line fit process from the relation,

$$\frac{d\tau}{dk} = \left(\frac{2\pi}{\tilde{\omega}+\omega_L} - T\right). \tag{7}$$

Using the calculated $A$ and $B$ as the initial guess, the *fit* function implemented in MATLAB is used to obtain the final estimation of $A$ and $B$. While using the *fit* function, we apply a filter on the root-mean-square error (RMSE) calculation to prevent unphysical fitting results due to interception of nearby dips originating from other nuclear spins. The filter is built based on the initial (A, B) values as follows:

$$filter(\tau) = \begin{cases} 1, & \tau \in [\tau_k - \sigma, \tau_k + \sigma] \\ 0, & \text{otherwise} \end{cases}. \qquad (8)$$

The fitting process is applied to all lines that were selected in the CPMG line fit method. However, some of the chosen lines can be based on mixed dips that come from several different nuclear spins with differing interaction strength. To remove the incorrectly selected lines, a post-selection process is applied to the fitted (A, B) pairs. The (A, B) pairs that come out from the fitting process are regarded as candidates for a good nuclear spin interaction representation. The Beam Search[39], a heuristic search strategy for memory saving based on the breadth-first search, is applied to find the optimal (A, B) pair configuration, which reconstructs the most similar signal compared to the given CPMG signal. The algorithm calculates the RMSE value with the filter for each (A, B) pair configuration similar to the fitting process.

**Results and Discussion**

We first build up our systematic approach with an example of virtually generated CPMG signals. We simulate the CPMG signal with randomly selected hyperfine interaction tensor components of ten individual nuclear spins within 2 nm radius from the NV center for a given $\pi$-pulse repetition number $N = 32$. Approximately 50000 random coordinates, which represent nuclear spins more than 2 nm away from the electron spin, in the diamond lattice having interaction strength of $A < 8$ kHz, $B < 0.25$ kHz were chosen to reproduce the spin bath signal (Fig. 2a). Figure 2b shows the Gaussian decomposition [33] of the input signal where the position, amplitude, and standard deviation value of each Gaussian are obtained using the algorithm described in the method section. Figure 2c shows the plot of the positions of the dips in the $\Delta\tau$ - $k$ fan diagram. As explained above, the automatic CPMG line fit method is applied to obtain the hyperfine interaction, which finds the best configuration of straight lines starting near origin covering most of the data points as shown in Fig. 2c, using solid lines. For each line found from the fit, $A$ and $B$ values are calculated by solving Eq. (4), and Eq. (6). These hyperfine parameters are used as initial guesses to fit each dip iteratively based on Eq. (2) to reach the final hyperfine interaction tensor components. The procedure is applied to each detected nuclear spin, and the accuracy of the fit is determined by comparing the reconstructed and input CPMG signal. As shown in Fig. 2d, the reconstructed CPMG signal, using the converged hyperfine interaction tensor components, shows excellent agreement with the input configuration of the spins. This shows less than 5% error compared to the values used in the signal generation. It also demonstrates the capability of the algorithm to automatically search the given nuclear spectra. The error of hyperfine interaction was calculated as,

$$Error = \frac{\sqrt{(A_{orig} - A_{obt})^2 + (B_{orig} - B_{obt})^2}}{\sqrt{A_{orig}^2 + B_{orig}^2}} \quad (9)$$

where $A_{orig}, B_{orig}$ are the ($A$, $B$) value of the original hyperfine interaction, and $A_{obt}, B_{obt}$ are the ($A$, $B$) value of the obtained hyperfine interaction, the result of our program.

We turn to discuss more systematic analyses of the performance of the developed algorithm. We first test the performance with numerically generated CPMG spectrums of a single nuclear spin with varying $A$ and $B$. Figure 3a shows the error of the hyperfine interaction strength from a single nuclear spin signal. The small $A$ domain ($|A| < 5$ kHz) was not detectable because of the overlap of the coherence dip signal with virtually generated spin bath and other nuclear spin signals with similar dip periods. As $A$ becomes larger in the large $A$ domain ($|A| > 70$ kHz), the slope of the line in $\Delta \tau - k$ plot becomes steeper. In the spin detection stage, this program considers only two additional clone layers ($M = 3$) of the point set while applying the CPMG line fit method to the $\Delta \tau - k$ plot to ensure a manageable computation time. As a result, the range of detectable slopes is limited, and a step jump of the error in the large $A$ domain is observed. Additionally, the small $B$ domain ($B < 5$ kHz) was not detectable because of the small size of the dips (smaller than the size of the threshold 0.05 given during the *Signal decomposition* process), and the large $B$ domain ($B > 120$ kHz) was hardly detectable because the shape of each of the dips deviates from a simple Lorentzian form to a more complex shape with several ripples around the dip typical for a CPMG dynamical decoupling sequence. We expect that more advanced techniques such as pattern analysis based on machine learning combined with numerical fits can enhance the detectability in this strongly coupled regime.

Since the algorithm is based on a nuclear spectrum, stemming from all nuclear spins interacting with the NV center together, it is important to determine the minimum resolution of the hyperfine coupling parameter that the algorithm can reliably distinguish. We analyze simulated CPMG signal, originating from two different spins to estimate the resolution of the dip frequency $f_p = \frac{\tilde{\omega} + \omega_L}{4\pi}$. One nuclear spin with fixed hyperfine interaction tensor components was used as the reference signal while the hyperfine coupling of the second nuclear spin was varied. Figure 3b and 3c show the error of the obtained hyperfine parameters of the second nuclear spin as a function of parameter variation from the reference spin. The results show that the main factor significantly affecting the resolution of the program is $A$ since it is directly related to the slope of each line in the spin detecting stage as well as the dip frequency of the nuclear spin. We empirically find that the obtained dip frequency resolution of the current algorithm is about 2 kHz, which showed an error less than 20%.

We further analyze the accuracy and applicability of this program by repeatedly testing the algorithm with virtual CPMG data sets containing multiple random nuclear spins, covering a wide range of $A$ and $B$. The values are based on possible lattice points in diamond and calculating the hyperfine interaction from a purely dipole-dipole coupling – neglecting the contact hyperfine and extended wavefunctions for simplicity – between the $^{13}$C nuclear spin and the electron spin of the NV center using,

$$H = -\frac{\mu_0 \gamma_e \gamma_n \hbar^2}{4\pi |\mathbf{r}|^3} (3(\mathbf{S}_e \cdot \hat{\mathbf{r}})(\mathbf{I}_n \cdot \hat{\mathbf{r}}) - \mathbf{S}_e \cdot \mathbf{I}_n) \qquad (10)$$

where $\mu_0 = 4\pi \times 10^{-7}$ H/m is the magnetic permeability of free space, $\gamma_e$ = 20824 MHz/T ($\gamma_n$ =10.708 MHz/T ) is the gyromagnetic ratio of the NV center electron ($^{13}$C nuclear ) spin, $\mathbf{S}_e$ is the vector of electron spin-1 operators, and $\mathbf{I}_n$ is the nuclear spin-1/2 operators vector. We confine the lattice points to be within 2 nm from the NV center, and consider spins having hyperfine interaction values within the range -100 kHz < $A$ < 100 kHz (5 kHz < $B$ < 100 kHz). Twenty lattice points in this range were randomly chosen to simulate the signal, and by repeating our algorithm, we record the error of the analysis result for each nuclear spin used to generate the signal.

Figure 3d shows the average error between the input hyperfine coefficients and the output of this method by accumulating 1000 configurations of 20 random spins. From this result, we find that the region of 80% confidence is 5 kHz < |A| < 70 kHz and 15 kHz < |B| < 80 kHz. The region of both |A| and |B| is constrained from the limitation of detecting a single nuclear spin signal as shown in Fig. 3a. In addition, for |B|, the overlap of the dips coming from different nuclear spins appears in the many spin signals, narrowing the allowed range of this parameter. The loss in confidence can be ascribed to two main factors; (1) the accuracy of the period of the dips, and (2) the accuracy of estimating the amplitude of each dip. The period of each nuclear spin signal is estimated using a large collection of dip points. This is generally robust to the presence of other nuclear spin signals, leading to frequency resolution 2 kHz (Fig. 3e). However, the amplitude of each dip can be disturbed when overlapped with other spin signals. As a result, the error of the hyperfine parameters mainly comes from the error in fitting the amplitude of the dip, affecting the accuracy of the perpendicular hyperfine interaction element $B$. As shown in Fig. 3b and c, when the dip frequency coming from two different nuclear spins is similar within 2 kHz, the dips coming from different spins interfere with each other. This brings in a non-linear distortion to the dip amplitude. It is found that among spins in the 80% confidence region, an average of 20% of the spins has another nuclear spin whose dip frequency difference is within 2 kHz, making it difficult to fit the signal from each nuclear spin independently (Fig. 3f). The effect of the interfering nuclear spin signal can be approximated from Eq. (2). When some error in $B$ comes in, represented as $\delta B$, the size of the dip is $P = \sin^2\left(\dfrac{N}{2}\dfrac{B}{\sqrt{(\omega_L+A)^2+B^2}}\right)$, and the error of the dip becomes $\dfrac{\delta P}{P} \sim 2\dfrac{\delta B}{B}$ up to the order of $O\left(\left(\dfrac{B}{\omega_L+A}\right)^4\right)$, for $A, B \ll \omega_L$. This result shows that the error in the amplitude of the dip $P$ directly affects the error of $B$. From this relation, it can be interpreted that if an additional spin signal with its dip size of 0.2 is overlapped with the original signal, the output $B$ value is expected to show 20% error.

We now show the application of the method to the experimental NV center-based nuclear spectroscopy data. The data was collected from a single NV center system at a low temperature with varying $\tau$ from 6 to longer than 50 μs[11]. The experimental data in ref. 11 were re-analyzed, and our algorithm reports 14 distinct nuclear spins (Fig. 4a). As an output of the algorithm, the hyperfine interaction tensor components of each spin were obtained. The simulated signal from these hyperfine parameters matched well with the experimental data, as shown in Fig. 4b. The obtained hyperfine interaction tensor components are given in Table 1. Based on the accuracy analysis performed above, spins 1, 2, 4, 10, 11, 12, and 14, marked with asterisk in the table, are in the reliable (*A, B*) region, and as expected, they show values less than 20% difference compared to the reported values from the experimental multidimensional spectroscopy on the same NV center and nuclear spin environment[23] (Table. 1).

This algorithm reported additional spins 3, 5, 6, 7, 8, 9 and 13, whose (*A, B*) values are out of the 80% confidence range. Spin 3 and 13 have their consistent match among the values of ref. 23, even though their *B* values were below the confidence threshold. Spin 5-9 show dip periods similar to the spin bath and overlap with each other as shown in Figure 4c. For longer $\tau$, the coherence dips for spin 5-9 become distinguishable from one another as shown in Fig. 4d. The reported hyperfine values of spin 5, 7, 8 and 9 show large deviations from all values in ref. 23, and spin 6 has a consistent match in ref. 23. While the algorithm treats each line/dip position group as coming from a single nuclear spin, due to limited spectral resolution of the CPMG data, it is highly likely that each spin label contains more than one nuclear spin. In the small *A* (< 5 kHz) regime, only the coherence dip period information, but not the individual *A* and *B* values, is reliable. Nevertheless, to explain the experimental coherence values significantly below 1 at the location of each of the dips coming from spin 5-9, at least one nuclear spin is necessary per coherence dip period. Overall, for the reported spins with (*A, B*) values out of the 80% confidence range, the dip period information of the spins can aid to set a spectral region to focus on when adapting more advanced spin spectroscopy methods with higher spectral resolution.

In this algorithm, the CPMG signal was simulated and analyzed based on Eqs. (1) and (2). For the purpose of extracting the hyperfine interaction coefficients *A* and *B*, these equations can be applied to spin-1/2 systems within dipolar hyperfine interaction approximation. Therefore, the application of this algorithm is not limited to the NV center-$^{13}$C coupling in the natural diamond environment which we demonstrated. We expect that applying this algorithm to different spin environments, such as in silicon carbide, would be possible.

In these equations, the interaction between the $^{13}$C nuclear spins was neglected because of the small interaction strength compared to the one between the NV center electron spin and the $^{13}$C nuclear spin. However, the nuclear-nuclear couplings can be significant, especially for larger $\tau$ and *N*, and can cause broadening or splitting of the dips[11,23]. This could impair the analytical performance of this program. Additional consideration of the interaction between $^{13}$C nuclear spins in the fitting stage will increase the capability of this program even for results obtained from the large time region.

Additionally, under the same nuclear spin environment, the CPMG signal of larger $N$ values shows more disordered results because the number of nuclear spins with detectable dips within the CPMG signal increase, elevating the probability of dip overlap between different nuclear spins in the resulting signal. Similar problems can arise in analyzing the CPMG signal obtained under an environment with $^{13}$C concentration higher than natural abundance. In this algorithm, it is assumed that a single straight line in the spin detection stage represents a single nuclear spin. However, it is possible that multiple nuclear spins with different hyperfine interaction tensor components share their dip frequency, and form a single line as implied in Figure 3b, and c. In this case, because the multiple spin signal is considered as a signal originating from a single nuclear spin in this algorithm, the fitting accuracy can be reduced. Since treating signals from a single straight line as a multiple spin signal takes more complex calculations and a longer time, it was not considered in this algorithm. Considering multiple nuclear spins in the fitting stage will increase the accuracy of the analysis.

These limitations bring in difficulty in analyzing CPMG signals coming from sequences with a large pulse repetition number. Nevertheless, the CPMG signals with large $N$ values are useful for detecting weakly coupled nuclear spins because the size of dips in the CPMG signal coming from a nuclear spin with small perpendicular hyperfine interaction strength becomes magnified, and the coherence time of the NV center increases. Further improvements in solving the non-linearity (for example, a combination of our numerical fit procedure with a more advanced machine learning approach, and including the nuclear spin-nuclear spin interactions to the model) should improve the general performance of our algorithm in the future.

**Conclusion**

In this study, we have proposed and demonstrated an automated algorithmic method to analyze the CPMG dynamical decoupling nuclear spectrum to detect $^{13}$C nuclear spins near the NV center in diamond. Our algorithm decomposes spectroscopy signals into Gaussians using the signal partitioning algorithm [33], detects nuclear spins automatically, and returns hyperfine interaction tensor components of each nuclear spin. We confirmed that our program successfully works for both simulated signals and an extended experimental data set. Moreover, we analyzed the range of applicability and limitations of this methodology, which can be improved with additional consideration of the interaction between nuclear spins, and signal processing regarding non-linearity due to multiple nuclear spins. With the possibility to be generalized to other spin-1/2 systems, our results show the first step of a systematic algorithmic approach, providing a useful and general tool for sensing complex spin structures, and furthermore, expanding quantum systems based on spin qubits.

**Data Availability**

All data generated within this study are available from the corresponding author on request.

**Author Contributions**

H. O. and J. Y. designed and implemented the program. K. J. and K. K. provided technical help. M. H. A. and T. H. T. provided theoretical background and experimental data. D. K. conceived and supervised the project. H. O, J. Y., and D. K. wrote the manuscript.

**Competing Interests**

The author(s) declare no competing interests.

**Acknowledgements**

This work was supported by the National Research Foundation of Korea (NRF) Grant funded by the Korean Government (MSIT) (No.2018R1A2A3075438, No.2019M3E4A1080144, No.2019M3E4A1080145) and the Creative-Pioneering Researchers Program through Seoul National University (SNU). This work was supported by the Netherlands Organisation for Scientific Research (NWO/OCW) through a Vidi grant, and as part of the Quantum Software Consortium programme (Project No. 024.003.037/3368). This project has received funding from the European Research Council (ERC) under the European Union's Horizon 2020 research and innovation programme (grant agreement No. 852410). This project (QIA) has received funding from the European Union's Horizon 2020 research and innovation programme under grant agreement No 820445.

**Figure Legends**

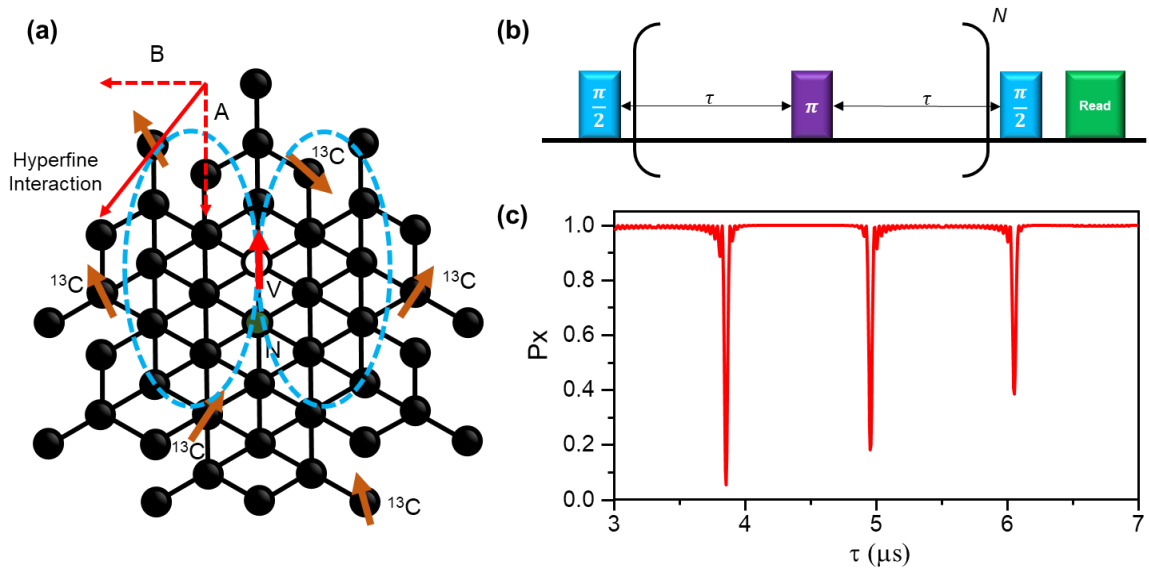

**Figure 1. Nitrogen vacancy (NV) center-based electron-nuclear spin register and CPMG pulse technique.** (a) Schematic diagram of the NV-center system. NV-center has spin quantum number $S = 1$ electron spin and interacts with nearby $^{13}C$ nuclear spins via the hyperfine coupling. Two states of the electron spin, $m_s = 0$ and $m_s = -1$, were chosen, and microwave pulses were resonant to these two states. (b) Schematic of Carr – Purcell – Meiboom – Gill (CPMG) dynamical decoupling pulse sequence. Each box indicates a single quantum gate that rotates the state vector around a specific axis in the Bloch sphere. (c) Example of CPMG spectroscopy data stemming from nearby single nuclear spin. $P_x$ represents the probability that the NV center spin state is preserved as a function of inter $\pi$- pulse duration $\tau$. The signal was simulated with a pulse repetition number $N = 32$, applied magnetic field $B_0 = 400$ G along the NV center axis, and hyperfine interaction tensor components $A = B = 50$ kHz.

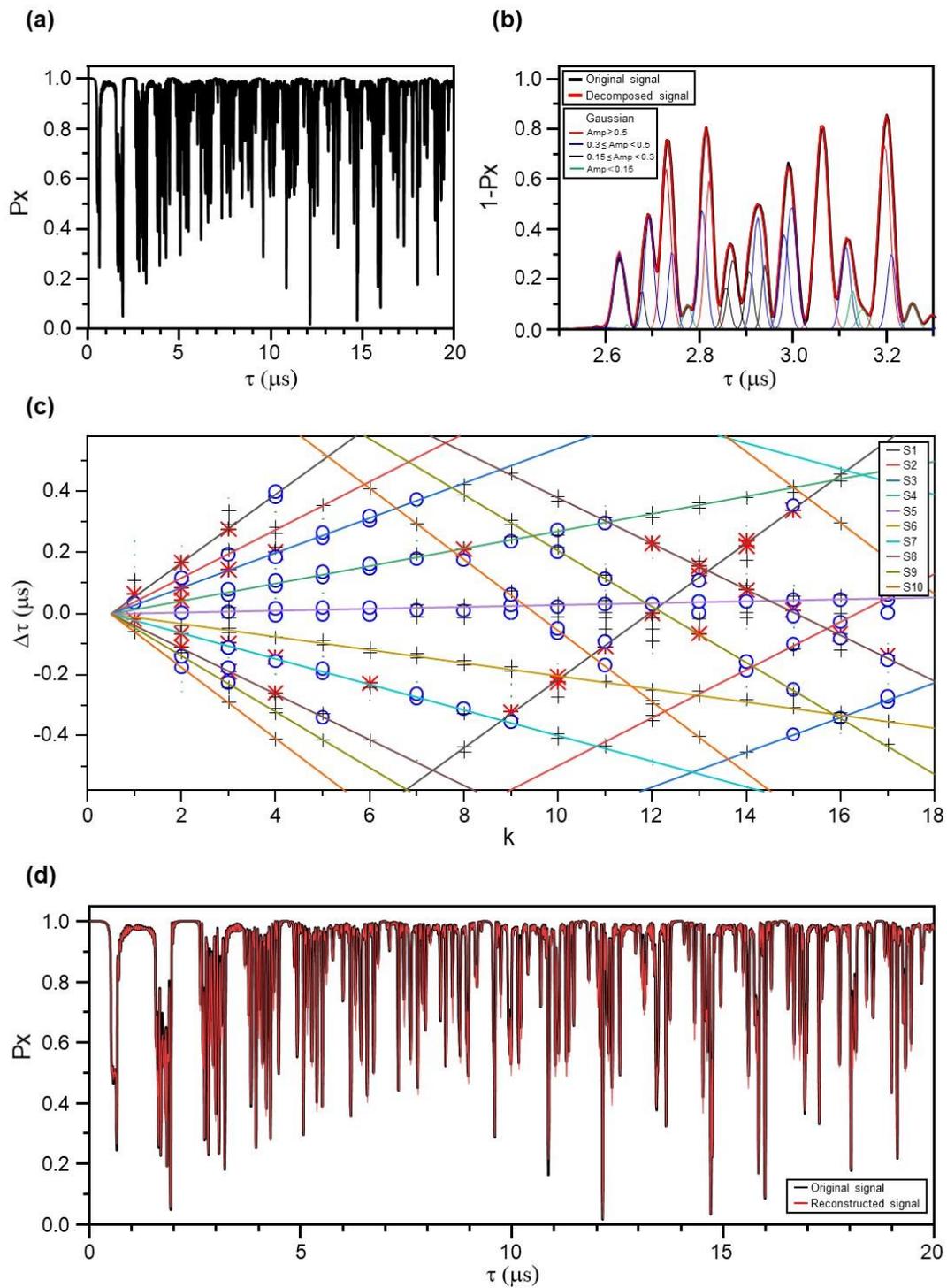

**Figure 2. Algorithm execution with virtual CPMG data.** (a) Simulated dynamical decoupling signal with ten nuclear spins. The hyperfine interaction tensor components of each nuclear spin were randomly generated. (b) Decomposition of the coherence dip with Gaussians by partitioning algorithm. Decomposed Gaussians are plotted with different colors depending on their amplitude. (c) The position of the coherence dips $\Delta\tau$ as a function of $k$. Each point

indicates a single dip extracted by Gaussian decomposition in the CPMG signal. The symbols represent the amplitude of the Gaussian used to fit each dip (For the Gaussian amplitude $a$, red asterisk: $a \geq 0.5$; blue circle: $0.3 \leq a < 0.5$; black cross: $0.15 \leq a < 0.3$; dots: $a < 0.15$). The colors of each marker correspond with the colors of decomposed Gaussians in (b). Lines are guides for the eye to group the points originating from the same nuclear spin. All ten nuclear spins used to generate the CPMG signal were detected. (d) Comparison of the generated data and the signal reconstructed from the detected spin parameters. The error of the hyperfine interaction tensor components of each nuclear spin is less than 5%.

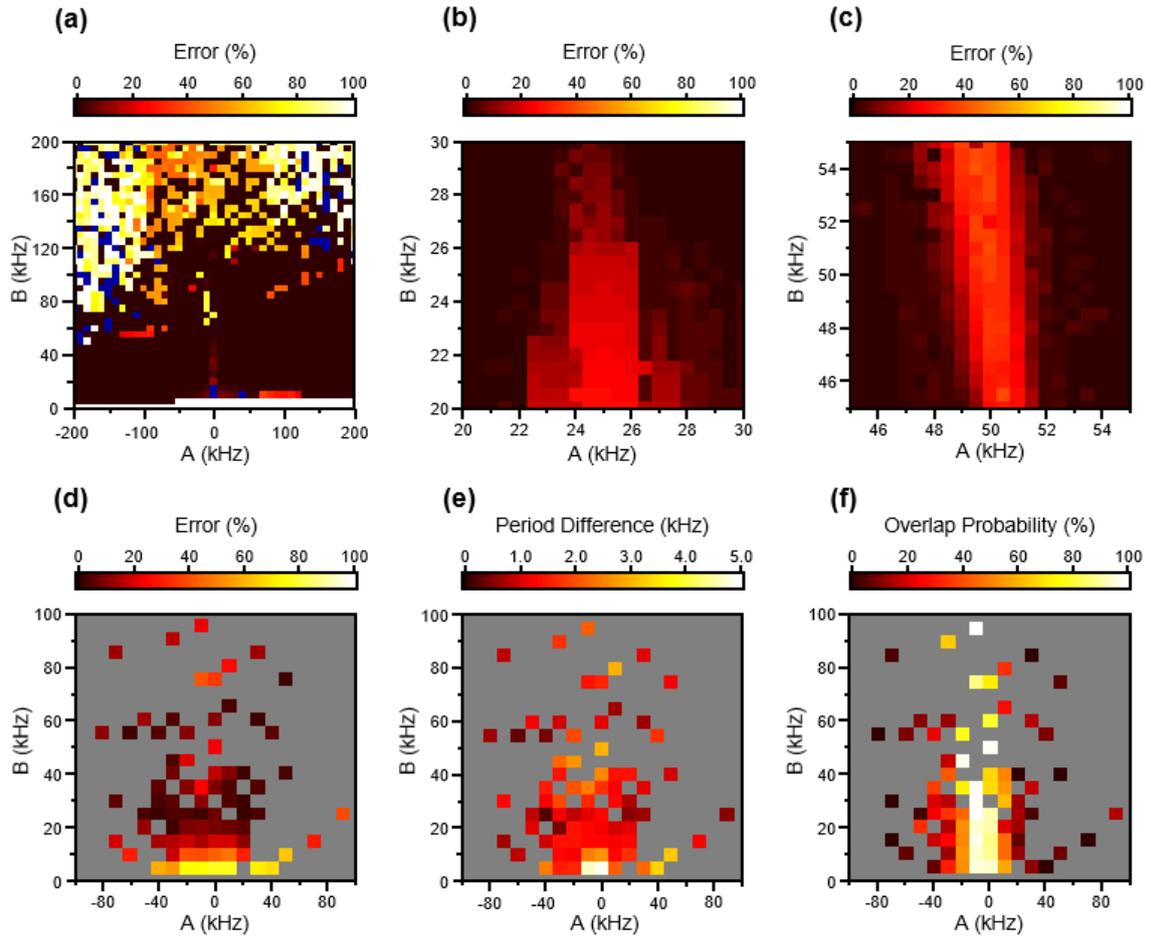

**Figure 3. Program performance analysis.** (a) Performance tests using signals originating from a single nuclear spin. The dark blue color indicates (*A*, *B*) values where the spin could not be detected (b-c). Spectroscopy resolution test using signals originating from 2 nuclear spins using the hyperfine parameters of the interference spin as (b) (25, 25) kHz, and (c) (50, 50) kHz. (d) Performance test using signals originating from randomly selected 20 spins. The color bar for (a-d) represents error for (*A*, *B*) values obtained by the algorithm. All errors were calculated using the distance on the *AB* plot between the original (*A*, *B*) used to produce the signal and the (*A*, *B*) obtained using this algorithm. The errors were averaged from the results of 1000 random configurations. (e) Averaged dip frequency difference between the original and obtained (*A*, *B*) values for all performance test results. (f) Probability of signal dip overlap for all spins used in the performance test. The gray color in the 2D map displayed in (d-f) represents the (*A*, *B*) values that cannot be achieved by the dipole-dipole interaction with nuclear spins in the diamond lattice. All signals were simulated with a pulse repetition number $N = 32$, applied magnetic field $B_0 = 400$ G. Pulse duration $\tau$ was varied from 0 to 50 μs with 5 ns interval.

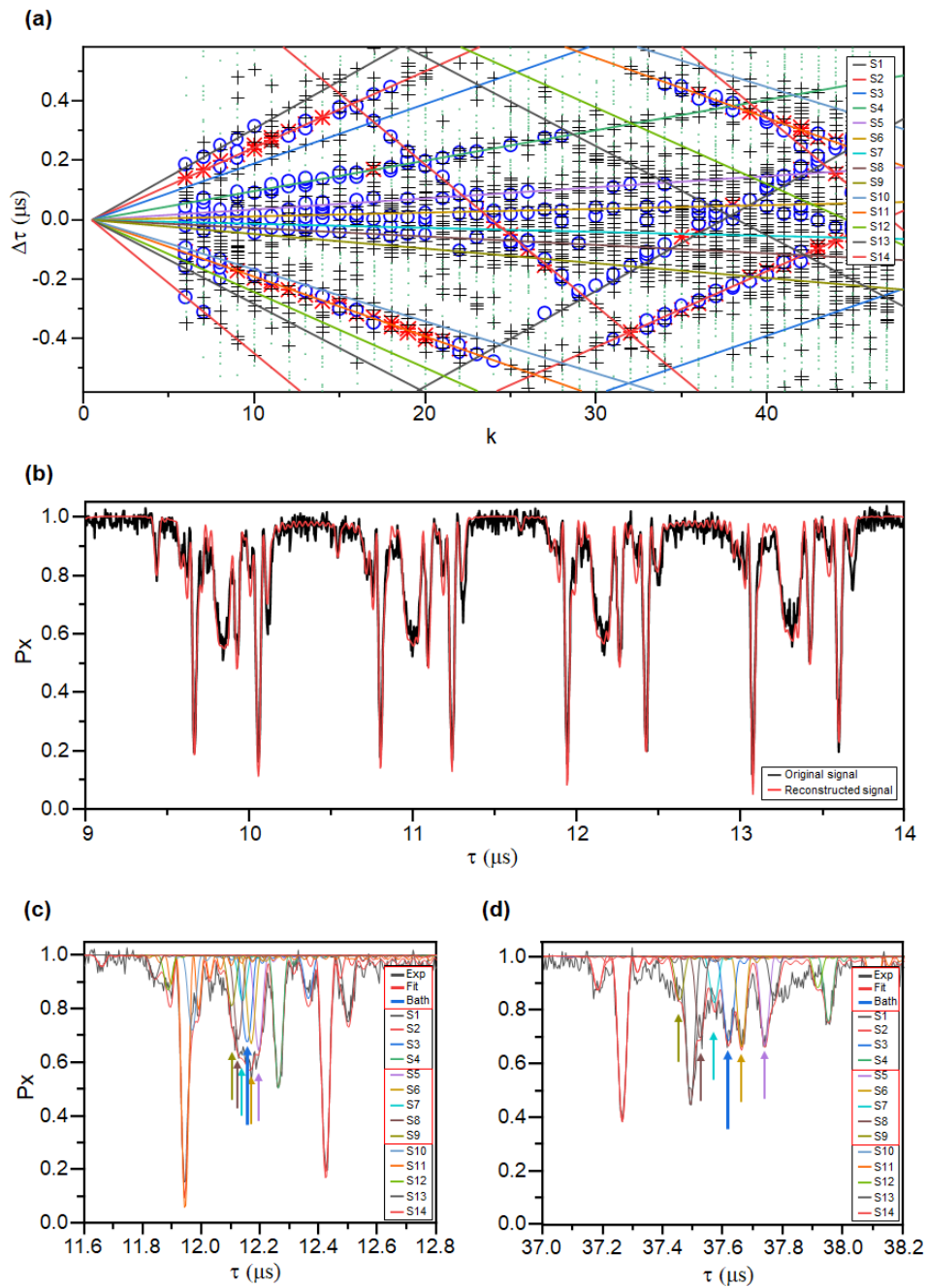

**Figure 4. Analysis result for experimental data.** (a) Detection of nuclear spins in the $\Delta\tau - k$ plot. Each point represents a single dip in the CPMG signal. Lines are guides for eye to group the points originating from the same nuclear spin. Points resting on a single line come from the same nuclear spin. Fourteen nuclear spins were detected in this data. The marker symbol follows the definition given in Fig. 2c. (b) Comparison of the experimental data and the signal reconstructed from the detected spin parameters. (c, d) Individual spin spectra from the reconstructed signal around (b) $\tau = 12$ µs and (c) $\tau = 38$ µs. The arrows mark the position of the dips coming from the spin bath and S5-9 in Table 1.

**Table 1 User-defined parameters used in the algorithm**

| Parameter | Value | Description |
|---|---|---|
| threshold | 0.05 | Threshold of the minimum amplitude of the output of the *Signal decomposition into Gaussians* process. Value matching the amplitude of noise is recommended to distinguish noise and signal. |
| $d_{max}$ | 1e-8 | Defines the maximum distance boundary from the line in which the dips are grouped during the *Detection of single nuclear spins* process. Should be defined related to the $\Delta\tau$ linewidth of an independent dip. |
| $M$ | 3 | Number of additional layers of dips used during the *Detection of single nuclear spins* process. Increasing $M$ expands the detectable $A$ range of the algorithm. |

**Table 2** Obtained hyperfine interaction tensor components between ¹³C and electron spin and reference values from experimental data[23]

| Spin | $A_{obt}$ [kHz] | $B_{obt}$ [kHz] | $A_{ref}$ [kHz] | $B_{ref}$ [kHz] |
|---|---|---|---|---|
| 1* | -23.7 | 17.82 | -24.399(1) | 24.81(4) |
| 2* | -20.58 | 39.80 | -20.569(1) | 41.51(3) |
| 3 | -14.83 | 14.30 | -14.548(3) | 10(1) |
| 4* | -8.10 | 22.12 | -8.029(1) | 21.0(4) |
| 5 | -3.14 | 18.66 | -2.690(4) | 11(1) |
| 6 | -1.25 | 16.59 | -1.212(5) | 13(1) |
| 7 | 0.87 | 13.12 | - | - |
| 8 | 1.83 | 17.10 | - | - |
| 9 | 3.52 | 14.09 | 3.618 | 0(2) |
| 10* | 11.30 | 59.87 | 11.346(2) | 59.21(3) |
| 11* | 13.04 | 16.51 | 14.07(2) | 13(1) |
| 12* | 19.30 | 15.89 | 20.72(1) | 12(1) |
| 13 | 22.67 | 14.86 | 23.22(1) | 13(1) |
| 14* | 36.07 | 29.67 | 36.308(1) | 26.62(4) |